\newif\ifprl
\prlfalse

\documentclass[prl,superscriptaddress,amsmath,amssymb,twocolumn]{revtex4-1}

\usepackage{amsthm}
\usepackage{graphicx}
\usepackage[all]{xy}
\usepackage[usenames,dvipsnames]{color}

\usepackage{environ}  

\ifprl
\newcommand{\mysection}[1]{{\bf #1} ---}
\NewEnviron{equ}{%
\(
  \BODY
\)
}
\else
\newcommand{\mysection}[1]{\section{#1}}
\NewEnviron{equ}{%
\begin{equation*}
  \BODY
\end{equation*}
}
\fi

\newcommand{\mc}[1]{\mathcal #1}

\newcommand{\ave}[1]{\langle #1 \rangle}
\newcommand{\dave}[1]{\langle\!\langle #1 \rangle\!\rangle}

\newcommand{\tr}{{\rm Tr}\,}
\newcommand{\one}{{\bf 1}}

\newcommand{\chan}[1]{\mathcal #1}

\newcommand{\be}{\begin{equation}}
\newcommand{\ee}{\end{equation}}
\newcommand{\bs}{\begin{split}}
\newcommand{\es}{\end{split}}

\theoremstyle{plain}

\theoremstyle{definition}

\begin{document}

\title{Quantum deconvolution}
\author{C\'edric B\'eny}
\address{Department of Applied Mathematics, Hanyang University (ERICA), 55 Hanyangdaehak-ro, Ansan, Gyeonggi-do, 426-791, Korea.}

\begin{abstract}
We propose a method for stably removing noise from measurements of a quantum many-body system. The question is cast to a linear inverse problem by using a quantum Fischer information metric as figure of merit. This requires the ability to compute the adjoint of the noise channel with respect to the metric, which can be done analytically when the metric is evaluated at a Gaussian (quasi-free) state. This approach can be applied effectively to $n$-point functions of a quantum field theory. For translation invariant noise, this yields a stable deconvolution method on the first moments of the field which differs from what one would obtain from a purely classical analysis. 
\end{abstract}

\maketitle

Measurements always come with some errors or noise. Given some characterization of the noise, we can attempt to reverse its effect as a post-processing step. However this has to be done with care~\cite{ferrie2012}, especially when the noise is strong, as the inverse operation may severely amplify statistical errors, leading to a useless output. 

If the process causing the noise is translation invariant (assuming we are dealing with a many-body system), then its effect on the signal is essentially a convolution. Hence the process of inverting its effect is called {\em deconvolution}. This is a problem with well established solutions at the classical level. 
It is used in microscopy, for instance, where the noise may come from optical imperfections.

In this work, we address the question of whether the knowledge that the system under observation is quantum rather than classical ought to inform the choice of reconstruction algorithm. We find that it does.

Rather than considering an algorithm that is itself quantum (a question related to quantum error correction), we focus here on classical post-processing: we assume that the noisy quantum state was first characterized classically, for instance via tomography, before attempting to remove the noise from the classical description. 

In our framework, the difference with the classical setting then stems from the special distinguishability structure of quantum states (quantum Fisher information metrics).

\mysection{General framework}
The setting and technique are closely related to those used in Refs.~\cite{beny2015a,beny2015b}. For simplicity, we will present the general approach in the context of finite-dimensional quantum theory. The resulting framework can be straightforwardly generalized to quantum field theory, which we do in an example below.

Suppose the true state of our system is characterized by a density matrix $\rho_t$, but we only have experimental access to its image $\mc N(\rho_t)$ under a quantum channel (completely positive trace-preserving map) $\mc N$. We consider a situation where we may produce many copies of that state so as to measure expectation values of observables. For the sake of the presentation, let us assume that we compile those measurements into an approximate tomographic description of $\mc N(\rho_t)$, which we call $\rho_m$. 
Below we show that the approach developed also works if we only know the expectation values of certain observables, such as those of field operators. 

In principle we could compute $\mc N^{-1}(\rho_m)$ since we have a classical description of $\mc N$. However, problems arise when $\mc N$ reduces the  distinguishability between certain states by a large factor. Small errors made in the tomography are then amplified accordingly by $\mc N$, leading to a garbled reconstruction. 

Let us analyse the problem in the neighbourhood of a state $\rho$, whose role will reduce classically to that of a Bayesian prior. We expand any smooth parameterization of the density matrices to first order in the parameters. This is equivalent to writing states as $\rho + \epsilon X$, where $X$ is self-adjoint and traceless, and expending any measure of distinguishability (divergence) $d(\cdot,\cdot)$ to lowest nontrivial order in $\epsilon$. If the measure is sufficiently smooth ($d$ could represent the relative entropy, or Bures distance for instance), we obtain 
\begin{equ}
d(\rho + \epsilon X, \rho + \epsilon Y) = \epsilon^2 \ave{X-Y,X-Y}_\rho + \mc O(\epsilon^4),
\end{equ}
where $\ave{\cdot,\cdot}_\rho$ is a real inner product (provided $\rho$ has no zero eigenvalue, otherwise it may be degenerate). Note that $X$ and $Y$ are really tangent vectors to the manifold of density matrices, and $\rho \mapsto \ave{\cdot,\cdot}_\rho$ is a Riemannian metric on that manifold. For a metric to make sense as a measure of distinguishability, it must contracts under the action of any channel: $\ave{\chan N(X),\chan N(X)}_{\chan N(\rho)} \le \ave{X,X}_\rho$. Such contractive Riemannian metrics have been characterized by Petz~\cite{petz1996}. They all take the form $\ave{X,Y}_\rho = \tr(X \Omega_{\rho}^{-1}(Y))$, where $\Omega_\rho$ is one of a many possible ``non-commutative multiplication by $\rho$''. Classically (when all operators commute), the only solution is $\Omega_\rho(A) = \rho A$, which defines the Fisher information metric.

The distinguishability between $\rho + \epsilon X$ and $\rho + \epsilon Y$ after the noise has been applied is given to lowest order in $\epsilon$ by 
\(
\epsilon^2 \ave{\mc N(X-Y),\mc N(X-Y)}_{\mc N(\rho)}. 
\)
One may think of this quantity as the contrast of the perturbation $X$ on the background $\rho$ in the presence of the noise $\mc N$. 

Now that we have a way of measuring how much information is lost about any given ``feature'' $X$, we can tackle the problem of stably inverting $\mc N$ on a state $\rho_t = \rho + \epsilon X$.

To do so, we need the adjoint (or rather {\em transpose} since the metric is real) $\mc N_*$ of $\mc N$, defined by the adjointness relation 
\begin{equation}
\label{transpose}
\ave{X, \mc N_* (Y)}_\rho = \ave{\mc N(X), Y}_{\mc N(\rho)}
\end{equation}
for all $X, Y$. Explicitely, $\mc N_*(X) = \Omega_\rho \circ \mc N^\dagger \circ \Omega_{\mc N(\rho)}^{-1}$, where  $\mc N^\dagger$ is the adjoint of $\mc N$ with respect to the Hilbert-Schmidt inner product: $\tr(E \mc N(X)) = \tr(\mc N^\dagger(E) X)$ for all operators $E$ and $X$.

Indeed, the extent to which the information about a normalized vector $X$ is lost is given by the norm of its image $\mc N(X)$, namely $\ave{\mc N(X), \mc N(X)}_{\mc N(\rho)} = \ave{X, \mc N_* \mc N(X)}_\rho$. The eigenvectors of the self-adjoint map $\mc N_* \mc N$ characterize the proper contraction directions. The corresponding eigenvalues---which are the singular values of $\mc N$ with respect to the metric---are between $0$ and $1$ and tell us how much information is preserved about the corresponding eigenvector.

For instance, for classical probability theory (diagonal density matrices), the only contractive metric is the Fisher information metric. The components of $\mc N_*$ are the conditional probabilities which arise from Bayes theorem applied to the conditional probabilities defining $\mc N$, with prior $\rho$.

The detailed approach that we use to approximately reverse the channel $\mc N$ then depends on the quality of our description of the noisy state, and hence how much amplification of its features we can tolerate.

For instance, if our measurements were infinitely accurate (which requires that we have access to infinitely many copies of the system, and that we can store infinite-precision numbers), then the optimal would be to multiply each eigenvector by the inverse of its eigenvalue if it is nonzero. This amounts to applying the pseudo-inverse $X = (\mc N_* \mc N)^{-1} \mc N_*(Y)$, yielding the reconstructed state $\rho + \epsilon X$ from the measured state $\rho_m = \rho + \epsilon Y$. Specifically, $X$ here minimizes the distance $\ave{\mc N(X)-Y,\mc N(X)-Y}_{\mc N(\rho)}$.

More realistically, one may use a map like $(\mc N_* \mc N + \Lambda)^{-1} \mc N_*$, where $\Lambda$ is some positive operator whose effect is to cutoff the amplification of vectors with small eigenvalue. Various such regularization schemes exist. 

At the other extreme, if we tolerate no amplification, then the best is to apply simply $\mc N_*$. This is what we will analyze in the examples below, as it gives us all the information that would be necessary to apply the other methods. Moreover, we will see that $\mc N_*$ alone can already provide some improvement of the data in relevant examples.

\mysection{Gaussian states and channels}
The effect of the adjoint map $\mc N_*$ can be computed when both $\mc N$ and $\rho$ are Gaussian (quasi-free). Gaussian states are characterised by a classical phase-space $V$ (possibly infinite-dimensional, but here we will use $V = \mathbb R^{2n}$ for simplicity) equipped with a symplectic form $\Delta$ specifying the Poisson bracket. One can always use coordinates such that $\Delta = \begin{pmatrix} 0 & 1 \\ -1 & 0\end{pmatrix} \otimes \one_n$. The algebra of operators is spanned by the unitary displacement operators (also called Weyl operators) $W_f$ ($f \in V$) such that $W_f W_g = e^{-\tfrac i 2 (f,\Delta g)} W_{f+g}$, where $(\cdot,\cdot)$ may denote, for instance, the canonical inner product on $\mathbb R^{2n}$. The unitary operators $W_f$ are projective representations of translations and ``boosts''. A state $\rho$ is entirely characterized by the expectation value $\rho(W_f)$ it assigns to $W_f$ (one may write $\rho(W_f) = \tr(\hat \rho W_f)$ where $\hat \rho$ is a density matrix). The state is Gaussian if
\begin{equ}
\rho(W_f) = e^{-\frac 1 2 (f,Af)},
\end{equ}
where $A$ is a positive semi-definite matrix (or operator on $V$ if $V$ is infinite-dimensional). In general one could substitute $f-f_0$ instead of $f$ in the exponent, where $f_0$ is some element of $V$. For simplicity we assume $f_0 = 0$.

The matrix $A$ is the {\em covariance} matrix. Indeed, let $\phi_f$ denote the canonical ``field'' operators such that $W_f = e^{i \phi_f}$. Then one can show that 
\begin{equ}
\rho(\phi_f \phi_g) = \frac{d^2}{ds dt} \rho(W_{tf} W_{sg})|_{s=t=0} = (f, (A+\tfrac i 2 \Delta)g). 
\end{equ}
This implies the constraint $A + \tfrac i 2 \Delta \ge 0$ (where the matrices are thought of as acting on a complexification of the phase space.)
In general, the expectation value of polynomials of order $n$ in the field operators can be obtained by $n$th-order derivatives of $W_f$ as above. 

A Gaussian channel is a completely positive trace-preserving map sending a Gaussian state with covariance matrix $A$ to another Gaussian state with covariance matrix 
\begin{equ}
B = X^\dagger A X + Y,
\end{equ}
where $X$ and $Y$ are matrices (or phase-space operators if $V$ is infinite-dimensional) satisfying 
\begin{equation}
\label{chanineq}
Y + \tfrac i 2 X^\dagger \Delta X + \tfrac i 2 \Delta \ge 0
\end{equation} 

For calculations, it is convenient to introduce the family of operators
\begin{equ}
G_f^M = W_f e^{- \frac 1 2 (\overline f, M f)},
\end{equ}
in terms of which the channel is characterised by
\begin{equation}
\label{Gaussianchan}
\mc N^\dagger(G_f^B) = G_{Xf}^A.
\end{equation}
If $A$ is the covariance matrix of the vacuum in a Fock space, then $G_f^A$ is the normal-ordering of the displacement operator $W_f$.
Here the dagger signifies the adjoint with respect to the Hilbert-Schmidt inner product: $\tr(X \mc N^\dagger(E)) = \tr(\mc N(X) E)$ for any matrices $X$ and $E$. In other words, we are working in the Heisenberg picture.

The coefficients of the Taylor expansion of $G_f^M$ at $f=0$, produce a complete basis of polynomials in the field operators $\phi_f$.
Let $\mc P_n^M$ be the subspace of operator spanned by the $n$-order derivatives of $G_f$ at $f=0$, which are polynomials of order $n$ in the operators $\phi_f$. It is clear from Eq.~\eqref{Gaussianchan} that $\mc N^\dagger$ maps $\mc P_n^B$ into $\mc P_n^A$. In fact, we show below that the converse property holds for $\mc N^\dagger_*$, which is central to making our approach practical for general metrics. But first, let us focus on a specific metric for which we can obtain a closed-form expression for $\mc N_*$.

\mysection{Square-root metric}
We consider the contractive metric given by
\begin{equation}
\label{metric1}
\Omega_\rho(X) = \sqrt \rho X \sqrt \rho,
\end{equation}
which we will refer to as the {\em square-root metric}.
The corresponding adjoint map $\mc N_*$ 
is {\em Petz' transpose channel}~\cite{petz1984}. Explicitely, $\mc N_*(X) = \rho^{\tfrac 1 2 }\mc N^\dagger(\mc N(\rho)^{-\tfrac 1 2}X \mc N(\rho)^{-\tfrac 1 2})\rho^{\tfrac 1 2 }$.

This map has the practical advantage of being a channel (completely positive and trace-preserving). Hence it is guaranteed to output a valid density matrix. Moreover, $\mc N_*$ has a relatively simple form if $\rho$ and $\mc N$ are Gaussian~\cite{Lami2017}. If $\rho$ is characterized by the covarience matrix $A$, and if $\mc N$ maps $A$ to 
\begin{equ}
B = X^\dagger A X + Y,
\end{equ}
then $\mc N_*^\dagger$ sends a Gaussian states with the covariance matrix $B'$ to a Gaussian state with the covarience matrix $X_*^\dagger B' X_* + (A - X_*^\dagger B X_*)$ where
\begin{equation}
\label{sol1}
X_* = R_{-\frac 1 2}^B (B+\frac i 2 \Delta)^{-1} X^\dagger (A + \frac i 2 \Delta) R^A_{\frac 1 2}.
\end{equation}
Here $\Delta$ is the symplectic form and $R_s$ is the linear operator on phase space which implements evolution by the imaginary time $is$, for the Hamiltonian $H$ such that $\rho = e^{-H}$. This is explained in more details in the 
\ifprl
Supplementary Material, 
\else
Appendix, 
\fi
where this equation is derived following Ref.~\cite{beny2015b}. 
An alternative form for $\mc N_*$ can be found in Ref.~\cite{Lami2017}.

For instance, for a harmonic oscillator characterised by the frequency $\omega$, mass $m=1$ at temperature $\beta$, we have 
\begin{equation}
\label{qho}
A = \frac 1 2 \coth(\beta \omega/2)\begin{pmatrix}1/\omega & 0 \\ 0& \omega \end{pmatrix}
\end{equation}
and 
$\Delta = \begin{pmatrix} 0 & 1 \\ -1 & 0\end{pmatrix}$,
then
\begin{equation}
\label{qhoi}
R_s^A = \begin{pmatrix} 
\cosh(\beta \omega_k s) & - i \omega_k \sinh(\beta \omega_k s) \\ i \sinh(\beta \omega_k s)/\omega_k & \cosh(\beta \omega_k s).
\end{pmatrix}
\end{equation}
The general case can be obtained by first decoupling the modes and then adapting the above formula.

The effect of this map is more explicitely characterised by
\begin{equation}
\label{solclosed}
\mc N_*^\dagger(G_f^A) = G_{X_* f}^B
\end{equation}
where the normal-odered displacement operators
\begin{equ}
G_f^M = W_f e^{- \frac 1 2 (\overline f, M f)}
\end{equ}
span the whole algebra of operators.  

We can obtain the action of $\mc N_*^\dagger$ on any polynomials in the field operators by differentiating Eq.~\eqref{solclosed} at $f=0$. For instance, a first order derivative in the direction $f$ yields
\begin{equ}
\mc N_*^\dagger(\phi_f) = \phi_{X_*f}. 
\end{equ}
We show below how this can be used in practice, even in a case where we only measured the expectation values of the field operators.

\mysection{Bures metric}
For other metrics, it may not be possible to obtain a general closed-form formula for $\mc N_*$. However, we can compute the action of $\mc N_*^\dagger$ on polynomials in $\mc P^A_n$ for any given $n$. The following can in principle be done for any of the contractive information metrics. For simplicity, we consider only the Bures metric defined by
\begin{equation}
\label{metric2}
\Omega_\rho(X) = \tfrac 1 2 (\rho X + X \rho).
\end{equation}
If we write $X = \Omega_\rho(G_f)$ and $Y = \Omega_\rho(G_g)$, we have
\begin{equation}
\begin{split}
\ave{X,Y}_\rho &= \tr(G_f^\dagger \Omega_\rho(G_g)) \\
&= \frac 1 2 \left({ e^{(f,(A+\tfrac i 2 \Delta) g)} +  e^{(f,(A-\tfrac i 2 \Delta) g)}}\right).
\end{split}
\end{equation}
(We extended the metric as $\ave{X,Y}_\rho = \tr(X^\dagger \Omega_\rho(Y))$ on non-self-adjoint operators).
Because the exponent on the right hand side is linear in $f$ and $g$, its derivatives are zero unless they are of the same order. This implies that the subspaces $\mc P_n^A$ and $\mc P_m^A$ are orthogonal in terms of the dual metric $\dave{A,B}_\rho = \tr(A \Omega_\rho(B))$ for $n \neq m$. 

Similarly, using Eq.\eqref{transpose}, we obtain
\begin{equ}
\dave{ G^B_f, \mc N_*^\dagger(G^A_g)}_{\mc N(\rho)} = \tfrac 1 2 \left({ e^{(Xf, (A+\tfrac i 2 \Delta) g)} + e^{(Xf, (A - \tfrac i 2 \Delta) g)} }\right).
\end{equ} 
This implies that $\mc N_*^\dagger$ is block diagonal in terms of the subspaces $\mc P_n^A$ and $\mc P_m^B$. 

Therefore, we can focus on determining the components of $\mc N_*^\dagger$ within each block. Let us consider $n=1$. The elements of both $\mc P_1^A = \mc P_1^B$ are simply the field operators $\phi_f$. Hence we know that there is an operator $X_*$ on phase space such that 
\begin{equ}
\mc N_*^\dagger(\phi_f) = \phi_{X_*f}.
\end{equ}
The adjointness relation $\dave{X,\mc N^\dagger(Y)}_{\rho} = \dave{\mc N_*^\dagger(X),Y}_{\mc N(\rho)}$ applied to $X = \phi_f$ and $Y = \phi_g$ yields
$
(f, A X g) = (X_* f, B g)$ for all $f,g \in V.
$
We conclude that
\begin{equation}
\label{sol2}
X_* = B^{-1} X^\dagger A = (X^\dagger A X + Y)^{-1} X^\dagger A.
\end{equation}

\mysection{Classical Fisher metric}
The classical statistical case can be obtained directly from Eq.~\ref{sol1} by using $\Delta = 0$ and $R_s^M = \one$ for any covariance matrix $M$. Hence $X_*$ simplifies to exactly the same formula as Eq.~\ref{sol2}.
This implies that the action of $\mc N_*^\dagger$ on the field operators in the classical case is identical to the quantum case with the Bures metric. However the two differ for higher order polynomials in the field operators.

\mysection{Reconstruction of first moments}
In practice these results can be used as follows. Consider for instance a discrete basis $e_i$ of the phase space $V$. Suppose we have measured the expectation values $F_i'$ of the observables $\phi_{e_i}$. Then, writing $\rho'$ for the quantum state after the action of $\mc N$, the reconstructed first moments, in the scheme where we only apply $\mc N_*$, are 
\begin{equation}
\begin{split}
F_i &= \tr(\mc N_*(\rho') \phi_{e_i}) = \tr(\rho' \mc N_*^\dagger(\phi_{e_i})) = \tr(\rho' \phi_{X_* e_i}) \\
&=\sum_j X_*^{ji} \tr(\rho' \phi_{e_j}) = \sum_j X_*^{ji} F_j',
\end{split}
\end{equation}
where $X_* e_i = \sum_j X_*^{ji} e_j$. In other word, if we think of the measured expectation values as forming an element of phase-space $F' \in V$, the reconstructed field is
\begin{equ}
F = X_*^T F'.
\end{equ}
Similary, since the action on first moments of $\mc N$ is given by $X$, then applying the reconstruction map $(\mc N_* \mc N + \Lambda)^{-1} \mc N_*$ amounts to $F = ((X_* X + L)^{-1}X_*)^T F'$, where $L$ is some positive operator representing the action of $\Lambda$ on field operators. That is, provided that $\Lambda$ is Gaussian.

\begin{figure}
\label{figex}
\includegraphics[width=0.9\columnwidth]{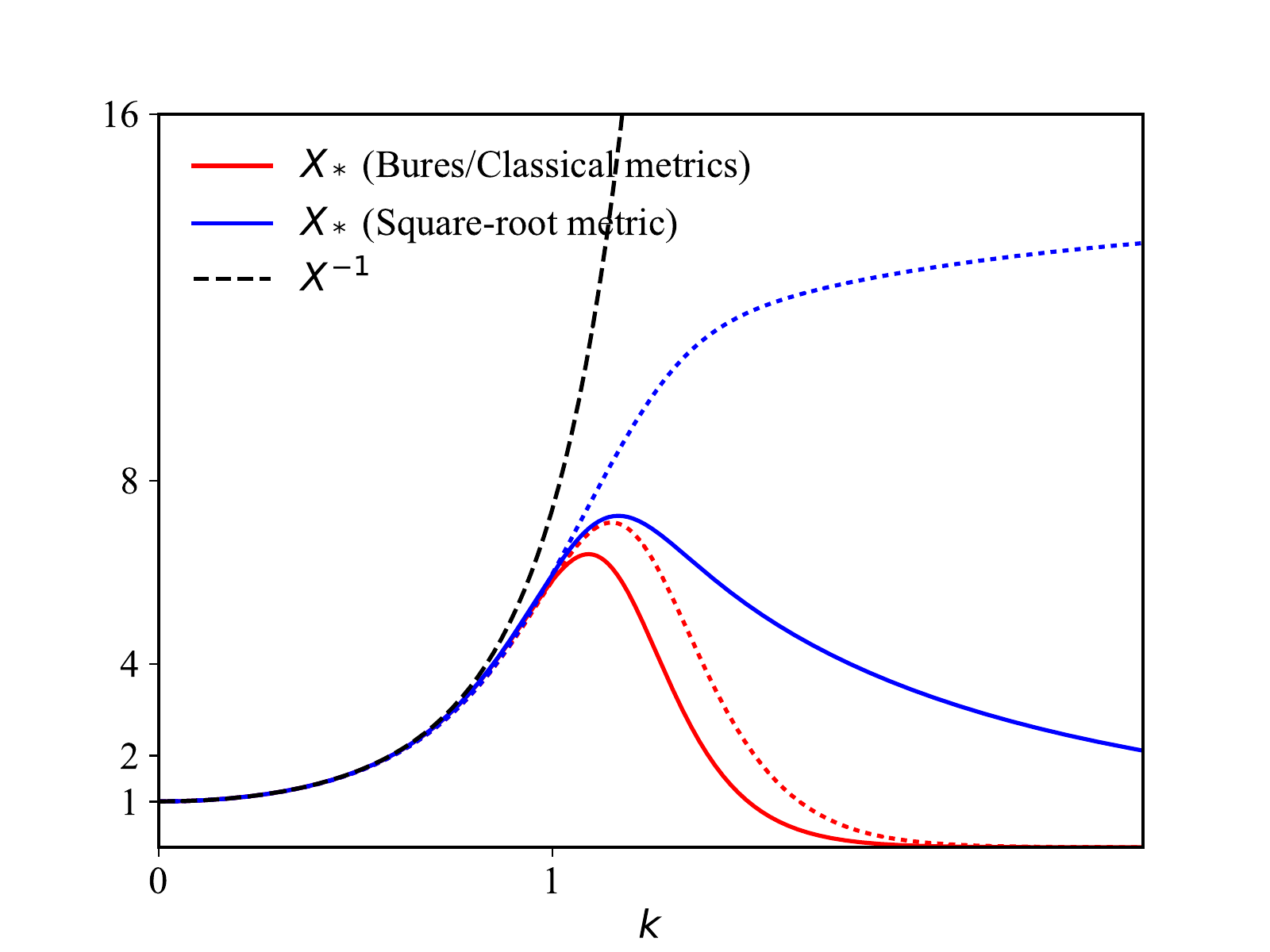}
\caption{
Spectrum of the adjoint reconstruction kernel $X_*$ for the example analyzed in the text. The spectrum is organized by the momentum magnitude $k$. For each value of $k$, $X_*$ has two eigenvalues corresponding to the canonical variable (solid line) and its conjugate (dotted line). The black dashed line represents the unregularized inverse of the convolution kernel: $X^{-1}$. The red line corresponds to the regularized reconstruction kernel $X_*$ computed for the Bures metric (Eq.~\eqref{sol2}) and is equivalent to the classical case using the Fisher informationm metric. The parameters $\beta = 0.01$, $\sigma = 2$ and $y_0 = 1$ were chosen to best illustrate the qualitatively different behaviour of $X_*$ computed from the square-root metric (Eq.~\eqref{sol1}).
}
\end{figure}

\mysection{Example}
As an example, we take $\rho$ to be a thermal state of a relativistic massless scalar field theory (as in Ref.~\cite{beny2015b}). Since the theory is translation-invariant, the covariance matrix is block-diagonal in terms of modes labelled by a momentum vector $k$. Each mode is characterised by a covariance matrix block $A_k$ given by the Eq.~\eqref{qho} with $\omega_k = |k|$. 

We use a channel that is translation covariant, so that $X$ and $Y$ are block-diagonal in terms of the modes $k$. Specifically, we use $X_k = e^{-\tfrac 1 2 \sigma^2 k^2} \one_2$ where $\one_2$ is the 2-by-2 identity matrix. This means that the effect of $\mc N$ on first moments (expectation values of the field operators) is a convolution with a Gaussian function of variance $\sigma$. Recall that the channel is also characterised by a second phase-space operator $Y$. Here we use $Y = \tfrac 1 2 y (\one - X)$, with $y \ge 1$. For $y = 1$, this saturates the constraints Eq.~\eqref{chanineq}. Here $Y$ can be thought of as imposing an uncertainty on the quadratures of each field modes. It must be nonzero so as to guarantee that the noisy state satisfies the Heisenberg uncertainty relations. 

The matrix $X_*$ is also block-diagonal in terms of the modes $k$, whether we use the metric given by Eq.~\eqref{metric1} or Eq.~\eqref{metric2}. In fact, $X_*$ is even diagonal in each block. Fig.~\ref{figex} illustrate a choice of parameters were both metrics lead to a qualitatively different type of regularization. 

Since the two metrics computed diverge from each other near pure states, it is to be expected that they give widely different results for a low temperature prior state $\rho$ ($\beta \gg 1$). 
The above example, however, is computed for a high-temperature prior. The reason for the strikingly different behaviour in this case can be traced to the fact that, because of the minimal choice $y = 1$, the coarse-grained modes of high momenta are nearly pure, in the sense that they saturate the Heisenberg uncertainty relations.

We observe that, in this particular situation, the square-root metric is more forgiving than the classical Fisher metric or the Bures metric, as it allows amplification of the high momentum modes. This is not true for all choices of parameters.

\mysection{Connection with other approaches}

The ``reconstruction kernel'' given by $X_*$ in Eq.~\eqref{sol2} alone yields the same formula as the standard {\em Wiener deconvolution} method~\cite{gonzalez04}. However, the statistical interpretation is different. Wiener deconvolution minimizes an {\em expected} distance under a noise model involving a deterministic convolution and an added random noise to the signal.
The eigenvalues of the operator $Y$ in that context then represents the power spectrum of the added noise. In particular, this implies that the reconstruction is interpreted as taking place on a sample of the signal, rather than on the average signal as done here. 

In the quantum setting, our approach is also related to the problem of finding a physical transformation (completely positive map, or channel) which reverses a channel on a given set of states. Indeed, a state does not lose relative entropy (a measure of distinguishability) with respect to a state $\rho$ precisely when $\mc N$ can be reversed on both states by a quantum channel. Moreover, the corresponding reverse map is the channel $\mc N_*$ defined from the square-root metric (Petz map)~\cite{petz1988}. Consequently this map is useful for quantum error correction~\cite{barnum2002}. This property has been generalized to approximate reversal in Refs.~\cite{fawzi2014,junge2016,sutter2016}. In comparison, our work presents a different statistical context where the reversal (reconstruction) map need not be completely positive. The connection can be made more explicit by noting that, for any of the contractive metric and with $\mc N_*$ the corresponding adjoint of $\mc N$ in that metric at the state $\rho$, then for any traceless $X$,
\begin{equ}
\|X\|_\rho^2 -\|\mc N(X)\|_{\mc N(\rho)}^2 \ge \|\mc N_* \mc N(X) - X\|_\rho^2
\end{equ}
where $\|X\|_\rho^2 = \ave{X,X}_\rho$.
This inequality is similar in spirit to that which is the subject of Refs.~\cite{fawzi2014,junge2016,sutter2016}.


\mysection{Acknowledgments}
This work was supported by the research fund of Hanyang University (HY-2016-2237).

\bibliography{../complete.bib}

\ifprl
\else

\newpage

\section*{Appendix}

Here we derive Eq.~\eqref{sol1}. We do the calculation in terms of the dual metric
\begin{equ}
\dave{E,F}_\rho = \tr(E^\dagger \Omega_\rho(F)),
\end{equ} 
which we extended to non-hermitian operators. 
With $\Omega_\rho(X) = \sqrt \rho X \sqrt \rho$, this is
\begin{equ}
\dave{E,F}_\rho = \rho(E^\dagger \rho^{\tfrac 1 2} F \rho^{-\tfrac 1 2 }).
\end{equ}
We observe that if we write $\rho$ as a Gibbs state for a quadratic Hamiltonian $H$, i.e., as $\rho \propto e^{-H}$, then
\begin{equ}
E \mapsto \rho^s E \rho^{-s} = e^{-sH} E\, e^{s H}.
\end{equ}
This is the imaginary time evolution generated by the Hamiltonian $H$. On a single canonical operator $\phi(f)$, we have $\rho^{s} \phi(f) \rho^{-s} = \phi(R_{s}^A f)$ where $R_{-it}^A$ is the linear phase-space evolution matrix corresponding to the quadratic Hamiltoian $H$ for time $t$. Here $A$ denotes the state's covariance matrix, which also uniquely defines $H$. 

The slight difficulty this introduces is that $R_{s}^A$ acts on a complexification of the phase space. We do not mean the usual technique of treating half of the phase space variables as imaginary, which would yield an $n$-dimensional complex vector space from a $2n$-dimensional real phase space. Instead, we simply allow all the vector coefficients to be complex, which yields a $2n$-dimensional complex vector space.

In terms of the normal-ordered displacement operators $G_f^A$, this is
\begin{equation*}
\rho^{\tfrac 1 2} G_{f}^A \rho^{-\tfrac 1 2} =  G_{R_{\frac 1 2}^A f}^A.
\end{equation*}
Therefore, the dual metric can be computed via
\begin{equation}
\label{app1}
\dave{G_f^A,G_g^A}_\rho = \rho((G_f^A)^{\dagger} G_{R_{\frac 1 2}^A g}^A).
\end{equation}

We need to be careful about how the displacement operators behaves on the complexified phase-space. We have
\begin{equ}
W_f^\dagger W_g = W_{g - \overline f} \, e^{\tfrac i 2 (f ,\Delta g)},
\end{equ}
where the phase-space scalar product is extended so as to be conjugate-symmetric. For instance, this implies
\begin{equ}
\rho(W_f) = e^{\tfrac 1 2 (\overline f, A f)}.
\end{equ}
where the overline denotes complex conjugation component-wise. 
It follows that 
\begin{equation}
\label{app2}
\rho((G_f^A)^\dagger G_g^A) = e^{(f, (A+ \tfrac i 2 \Delta) g)}.
\end{equation}

The adjointness relation defining $\mc N_*$ reads
\begin{equation}
\label{adjdual}
\dave{G_f^A,\mc N^\dagger(G_g^B)}_\rho = \dave{\mc N^\dagger_*(G_f^A),G_g^B}_{\mc N(\rho)}
\end{equation}
where $A$ is the covariance matrix for $\rho$ and $B$ is the covariance matrix for $\mc N(\rho)$, namely $B = X^\dagger A X + Y$. 
Using
\begin{equ}
\mc N^\dagger(G_g^B) = G_{Xg}^A, 
\end{equ} 
and Eq.~\eqref{app2}, we obtain 
\begin{equation}
\label{app3}
\dave{G_f^A,\mc N^\dagger(G_g^B)}_\rho = e^{(f, (A+ \tfrac i 2 \Delta) R_{\frac 1 2}^A Xg)}
\end{equation}
Hence, with the ansatz
\begin{equ}
\mc N_*^\dagger(G_f^A) = G_{X_*f}^B,
\end{equ}
Eq.~\eqref{adjdual} is satisfied if
\begin{equation}
\label{app4}
e^{(f, (A+ \tfrac i 2 \Delta) R_{\frac 1 2}^A Xg)} = e^{(X_* f, (B+ \tfrac i 2 \Delta) R_{\frac 1 2}^B g)}
\end{equation}
for all $f$, $g$,
which implies
\begin{equ}
(A+ \tfrac i 2 \Delta) R_{\frac 1 2}^A X = X_*^\dagger (B+ \tfrac i 2 \Delta) R_{\frac 1 2}^B,
\end{equ}
or
\begin{equ}
X_* =  (B+ \tfrac i 2 \Delta)^{-1} (R_{-\frac 1 2}^B)^\dagger X^\dagger (R_{\frac 1 2}^A)^\dagger (A+ \tfrac i 2 \Delta).
\end{equ}
To obtain the form of Eq.~\eqref{sol1}, we use the fact that $(R^A_s)^{-1} = R^A_{-s} = \overline{R^A_s}$, which can be seen for instance from the explicit expression Eq.~\eqref{qhoi}. We also use the fact that the covariance matrix and symplectic structure are invariant under $R_s$, i.e., such that that $(R^A_s)^T(A+ \tfrac i 2 \Delta ) R^A_s = (A+ \tfrac i 2 \Delta )$. From these properties it follows that
\begin{equ}
(A+ \tfrac i 2 \Delta ) R^A_s = (R^A_s)^\dagger (A+ \tfrac i 2 \Delta ),
\end{equ}
which yields
\begin{equ}
X_* = R_{-\frac 1 2}^B (B+ \tfrac i 2 \Delta)^{-1} X^\dagger (A+ \tfrac i 2 \Delta) R_{\frac 1 2}^A.
\end{equ}

\fi
 
\end{document}